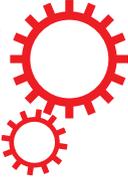

# Spin Seebeck effect and thermoelectric phenomena in superconducting hybrids with magnetic textures or spin-orbit coupling

Marianne Etzelmüller Bathen[1,2] & Jacob Linder[1]

We theoretically consider the spin Seebeck effect, the charge Seebeck coefficient, and the thermoelectric figure of merit in superconducting hybrid structures including either magnetic textures or intrinsic spin-orbit coupling. We demonstrate that large magnitudes for all these quantities are obtainable in Josephson-based systems with either zero or a small externally applied magnetic field. This provides an alternative to the thermoelectric effects generated in high-field (~1T) superconducting hybrid systems, which were recently experimentally demonstrated. The systems studied contain either conical ferromagnets, spin-active interfaces, or spin-orbit coupling. We present a framework for calculating the linear thermoelectric response for both spin and charge of a system upon applying temperature and voltage gradients based on quasiclassical theory which allows for arbitrary spin-dependent textures and fields to be conveniently incorporated.

Current device technology utilizing the electronic charge degree of freedom is rapidly approaching the limit of realizable computational power. The field of spintronics, which aims to incorporate the electron spin degree of freedom into devices with novel functionalities, has emerged as a promising alternative to silicon-based transistor technology[1]. Among the spin-dependent effects already incorporated into modern device technology are the spin-transfer torque (STT)[2] and the giant magnetoresistance (GMR)[3], which are used for memory applications. The key quantity to control for a wider range of application areas to emerge is how long a particle remains in one spin state, as spin coherence and control are essential for efficient and reliable operation of spintronic devices. Superconducting materials have attracted great deal of attention in this respect, as superconducting order increases the electron spin-flip relaxation time compared to the normal non-superconducting state[4–8]. Moreover, hybrid systems composed of superconductors and materials with properties such as textured magnetism and spin-orbit coupling contain the capability of generating spin-polarized supercurrents[9–18]. These and related properties of superconducting systems have caused the emergence of the field known as superconducting spintronics[19]. The superconducting order considered herein complies with the Bardeen-Cooper-Schrieffer (BCS) theory of superconductivity[20], where lattice vibrations cause two electrons of opposite spins to attract each other in contrast to the usual repulsive Coulomb interaction. This particle attraction causes the formation of zero-spin singlet Cooper pairs, and is the cause of conventional superconductivity[21].

Thermoelectric effects is the common denominator for the Seebeck effect and the opposite Peltier effect[22–24], and involve the generation of charge or heat currents upon applying a temperature or voltage bias. Superconductors have traditionally been regarded as poor hosts for thermoelectric effects and incapable of efficiently converting thermal energy into electric currents and vice versa. However, over the last few years, the combination of superconductivity and magnetism has challenged this notion, after very large thermoelectric tunneling currents were predicted in superconductor/ferromagnet (S/F) hybrid structures[25–27]. The prediction of thermoelectric effects comparable to those attainable in the best bulk thermoelectric semiconductors[27] being present in S/F systems exposed to strong external magnetic fields was recently experimentally verified[28]. Employing superconducting bilayers where both superconductors are exposed to strong external magnetic fields instead, resulting in Zeeman-split superconductors, was recently reported to further enhance these effects significantly[29]. Electron cooling in superconducting spin-filter junctions[30,31] and thermoelectric effects in superconducting quantum dot systems[32] have also been studied.

[1]Department of Physics, NTNU, Norwegian University of Science and Technology, N-7491 Trondheim, Norway. [2]Department of Physics, Centre for Materials Science and Nanotechnology, University of Oslo, N-0316 Oslo, Norway. Correspondence and requests for materials should be addressed to J.L. (email: jacob.linder@ntnu.no)

The thermoelectric phenomena in question include both electronic currents generated by a temperature bias, heat currents generated by a voltage difference and pure spin currents induced by a temperature gradient applied across the device. Spin currents of this kind are not dependent on the presence of spin-polarized superconductor/ferromagnet interfaces, provided that there is a spin-dependent particle-hole asymmetry on at least one side of the barrier interface. In the case of the S/F hybrids, this is achieved by the Zeeman-splitting of the superconducting density of states induced by an external magnetic field. Comparisons to commonly known thermoelectrics can be made using the Seebeck coefficient $\mathcal{S}$ and the thermoelectric figure of merit $ZT$. The thermoelectric materials currently available are capable of achieving $ZT \simeq 2$ and $\mathcal{S} \sim 1\,\mathrm{mV/K}$[33], which is rivaled by the superconducting bilayers. Consequently, thermoelectric superconducting hybrids provide a promising alternative in several low-temperature thermoelectric application areas, such as electron refrigeration and very precise thermal sensing.

The disadvantage to using Zeeman-split superconducting hybrids for this purpose resides within the necessity of applying strong magnetic fields on the order[28,34] of ~1 T for controllable thermoelectric effects to arise. This presents a significant challenge when considering potential application areas for superconducting thermoelectric devices. Therefore, this work will focus on expanding the study of thermoelectric superconducting hybrids to material systems where large applied magnetic fields are not needed. In the Zeeman-split S/F bilayers, the magnetic fields impose a spin-dependent asymmetry on the superconducting density of states, allowing the amount of particles residing in each spin state tunneling through the insulating barrier between the materials to be uneven. Spin-polarized tunneling currents and pure spin currents driven by applied voltage and temperature biases are the predicted result. The material systems studied within this work must replace the spin-splitting effect of the large external magnetic fields to enable thermoelectric phenomena. The material properties capable of imposing spin-splitting effects on the superconducting density of states studied herein include spatially varying ferromagnetism and spin-orbit coupling, neither of which depend on large external fields to achieve the desired results. The effect of intrinsic spin-orbit interactions has recently been shown to lead to interesting quantum transport phenomena in diffusive superconducting structures[35–41].

Similarly to ref. 42 we will consider a Josephson-based geometry which allows for an additional control parameter in the form of the superconducting phase difference $\Delta\theta$ across the junction[43,44]. Josephson junctions consist of a non-superconducting material placed between two superconducting reservoirs. The latter are assumed to be large when compared to the central component so that they may be treated as bulk BCS superconductors. Within this work, the central material is a semiconducting, metallic or ferromagnetic nanowire, which is separated from the superconductors via interfaces with low transparency for particle transport. Superconducting Cooper pairs may cross the tunneling barrier into the central material through a process known as the Holm-Meissner or proximity effect occurring between materials grown together in good contact[45]. Superconducting order can exist throughout the nanowire depending on the distance from the interface, magnetic order and the superconducting phase difference in the case of Josephson junctions. The inverse proximity effect is the influence of the other electronic system on the superconductor. This can affect both the superconducting critical temperature and the superconducting energy gap parameter, or induce e.g. ferromagnetic order within the superconductor[46]. The inverse proximity effect has a negligible impact, and can be disregarded, if the superconductor is very large compared to the adjacent material and interface transparency is low[10]. The thermoelectric phenomena considered herein depend on what is known as the triplet proximity effect, where magnetic texturing adjacent to the superconductor causes spin mixing and spin rotation of the singlet Cooper pairs, converting them into triplet Cooper pairs which can be spin-polarized.

The mathematical framework used in previous literature to predict thermoelectric effects arising in superconducting hybrids assumes collinear spin polarization, i.e. magnetic fields and materials are polarized along only one axis. When incorporating magnetic texturing and spin-orbit coupling, the arbitrary orientation of the spin-dependent fields existing in the systems must be taken into account. Within this paper, we extend the mathematical framework to encompassing materials with arbitrary magnetic texturing. For this purpose a quasiclassical approach based on the Keldysh Green function formalism will be employed, in a similar manner as in ref. 25, but here extended from collinear magnetic alignment and including a computation of the spin Seebeck effect. The thermal generation of a spin current and an associated spin voltage is known as the spin[47] or spin-dependent[48] Seebeck effect (we will stick with the former notation in this manuscript). Within the quasiclassical approximation, only particles with energies close to the Fermi surface are assumed to contribute to transport, and the Green function matrices are nearly isotropic with respect to momentum[49]. The second assumption is valid in highly diffusive systems where impurity scattering is dominant and extinguishes the anisotropic part of system dynamics[50].

## Theory

The geometry considered is that of a normal metal electrode (N) coupled to a nanowire (X) connecting two superconducting reservoirs (S) and forming an S/X/S Josephson junction. The electrode and nanowire are connected via a ferromagnetic or non-polarized insulator, as shown in the top panel of Figs 1, 2 and 3. The nanowire is either a conical ferromagnet, a normal metal with spin-active interfaces to the superconductors, or a spin-orbit coupled semiconductor. The central nanowires impose the necessary spin-splitting on the superconducting density of states, causing only low or no external magnetic fields to be necessary for thermoelectric tunneling currents to arise between the electrode and the nanowire. Thermoelectric phenomena occur as a result of quasiparticle tunneling from the nanowire to the electrode, with the quasiparticles having different tunneling probabilities depending on their spin state and the polarization of the interface. The charge, heat and direction-dependent spin currents across the tunneling barrier between the normal metal electrode and the central nanowire of the Josephson junction are defined by

$$I_q = \frac{N_0 e D A}{4} \int_{-\infty}^{\infty} dE \; \mathrm{Tr}\{\hat{\rho}_3 [\check{g}_R(\partial_x \check{g}_R)]^K\}$$
$$\dot{Q} = \frac{N_0 D A}{4} \int_{-\infty}^{\infty} dE E \; \mathrm{Tr}\{[\check{g}_L(\partial_x \check{g}_L)]^K\}$$
$$I_s^\nu = \frac{N_0 \hbar D A}{8} \int_{-\infty}^{\infty} dE \; \mathrm{Tr}\{\hat{\rho}_3 \hat{\tau}_\nu [\check{g}_R(\partial_x \check{g}_R)]^K\} \quad (1)$$

within the quasiclassical framework, where the $8 \times 8$ Green function matrices $\check{g}$ are propagators for the particle and hole states and contain the information necessary for describing particle dynamics within the system. $N_0$ is the Fermi level density of states, $A$ is the interface contact area, $D$ is the diffusion coefficient, $e$ is the electronic charge and $\hbar$ is Planck's reduced constant. The charge and spin currents are defined as those flowing on the right side of the junction, in the normal metal electrode, while the heat current is defined as flowing from the nanowire to the electrode. $\hat{\tau}_\nu = \mathrm{diag}(\sigma_\nu, \sigma_\nu^*)$ is the $4 \times 4$ Pauli matrix in Nambu space in each spatial direction $\nu = \{x, y, z\}$, and $\hat{\rho}_3 = \mathrm{diag}(\underline{1}, -\underline{1})$ is the Nambu space generalization of the **z**-aligned spin space Pauli matrix. $\underline{1}$ is the $2 \times 2$ unity matrix. $E$ is the quasiparticle energy in relation to the Fermi level, $L$ ($R$) denotes left (right) of the interface and $\check{g}$ is the $8 \times 8$ Green function matrix in Keldysh space[49–53]:

$$\check{g} = \begin{pmatrix} \hat{g}^R & \hat{g}^K \\ 0 & \hat{g}^A \end{pmatrix}. \quad (2)$$

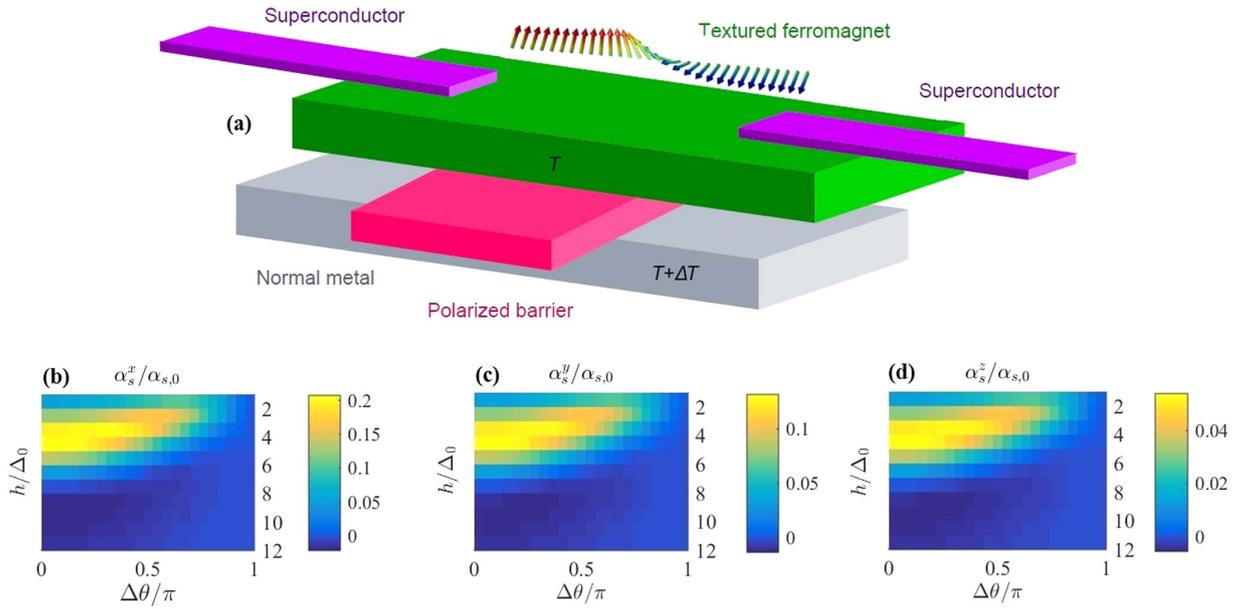

**Figure 1. Setup and thermoelectric effects for a superconducting structure incorporating a conical ferromagnet.** (**a**) Schematic of the proposed setup for observation of thermoelectric effects. Tunneling occurs from the center of a ferromagnetic nanowire into a normal metal electrode through an insulating barrier. The nanowire is connected to two superconducting reservoirs via low transparency interfaces. The magnetic structure of the nanowire is that of the conical ferromagnet holmium (Ho). *Bottom panel*: Thermal spin coefficient (**b**) $\alpha_s^x$, (**c**) $\alpha_s^y$ and (**d**) $\alpha_s^z$. The polarization of the barrier separating the normal metal and the textured ferromagnet is $P = 0$, and $\alpha_{s,0} = G_T \hbar \Delta_0 / e^2$.

We assume steady-state conditions in order to remove the time parameter from the equations of motion for the system, along with constant temperature and local equilibrium on each side of the junction. The chemical potential to the left of the barrier is defined as $\mu_L = 0$ for reference and the chemical potential on the right as $\mu_R = eV_R$. The Green function matrices in $4 \times 4$ Nambu space are expressed in terms of each other as

$$\hat{g}^A = -\hat{\rho}_3 \hat{g}^{R\dagger} \hat{\rho}_3, \quad \hat{g}^K = \hat{g}^R \hat{h} - \hat{h} \hat{g}^A, \quad (3)$$

where $\hat{h}$ is the non-equilibrium distribution function matrix

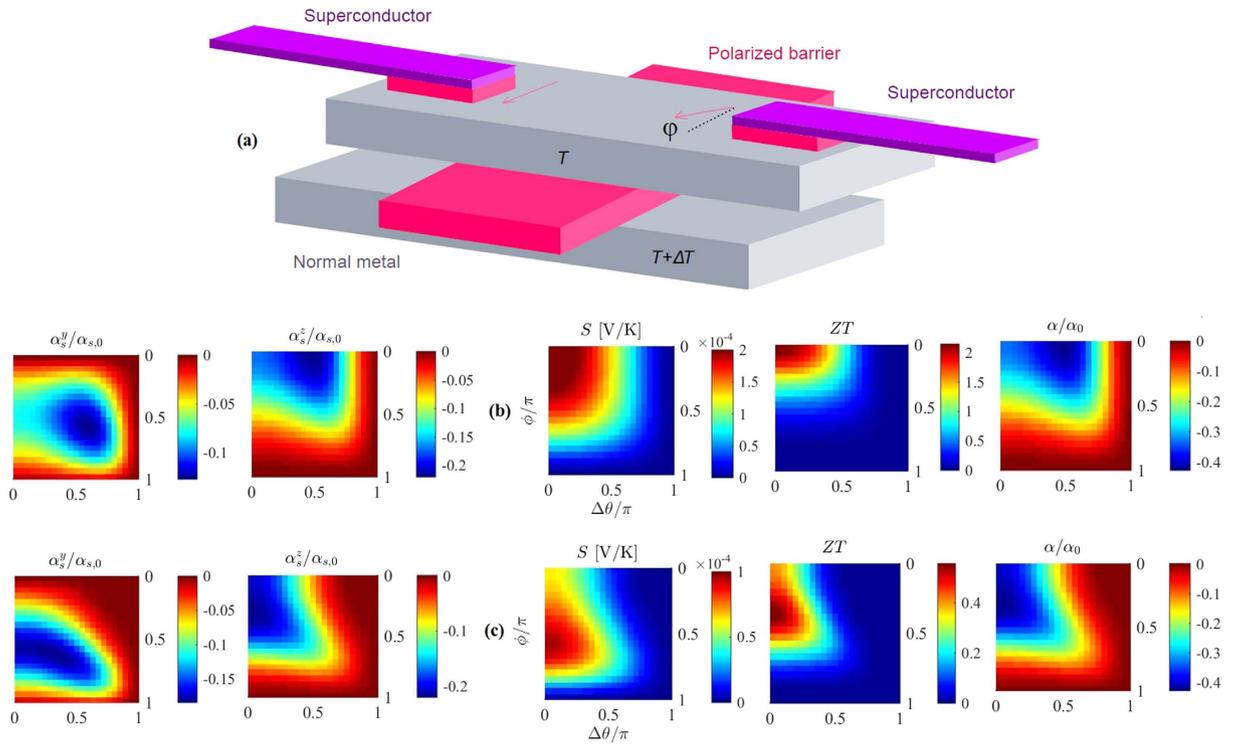

**Figure 2. Setup and thermoelectric effects for a superconducting structure incorporating spin-active interfaces.** (**a**) Schematic of the proposed setup for observation of thermoelectric effects. Tunneling occurs from the center of a normal metal nanowire into a normal metal electrode via a ferromagnetic insulator. The nanowire is connected to two superconducting reservoirs via weakly polarized tunneling barriers. One magnetic S/N interface is aligned along the **z**-axis while the magnetization direction of the other can be varied within the **yz**-plane. *Left panels*: Spin thermal coefficients $\alpha_s^y/\alpha_{s,0}$ (left column) and $\alpha_s^z/\alpha_{s,0}$ (right column). The polarization for nanowire/electrode tunneling is $P = 0$ and the polarization for S/N tunneling is encompassed by $G_{MR} = 0.1$. Spin-dependent phase shifts due to scattering at the S/N interfaces are governed by (**b**) $G_\varphi = 0.5$ in the top row and (**c**) $G_\varphi = 1.05$ in the bottom row. *Right panels*: Seebeck coefficient (left column), thermoelectric figure of merit (middle column), and thermoelectric coefficient $\alpha/\alpha_0$ (right column), where $\alpha_0 = G_T \Delta_0/e$. The polarization of the ferromagnetic insulator separating the nanowire and the electrode is $P = 97\%$, while the polarization for S/N tunneling is included in $G_{MR} = 0.1$.

$$\hat{h}_j = \begin{pmatrix} \tanh\left(\dfrac{\beta_R}{2}(E+\mu_j)\right)\mathbf{1} & \mathbf{0} \\ \mathbf{0} & \tanh\left(\dfrac{\beta_R}{2}(E-\mu_j)\right)\mathbf{1} \end{pmatrix} \tag{4}$$

under the conditions described above. Combining Equations (3) and (4), the Keldysh Green function matrix on the left side of the barrier takes the form

$$\hat{g}^K = \tanh\left(\frac{\beta E}{2}\right)(\hat{g}^R - \hat{g}^A), \tag{5}$$

where $\beta = 1/k_B T$. The current expressions defined in Equation 1 can be expanded using Eschrig's boundary conditions for arbitrarily polarized interfaces defined in ref. 54:

$$\check{g}_{R(L)}\partial_x \check{g}_{R(L)} = \pm \frac{1}{4e^2 N_0 DA}\left[G_0 \check{g}_L + G_{MR}\check{g}_L + G_1 \check{\kappa}\check{g}_L\check{\kappa} - iG_\varphi \check{\kappa}, \check{g}_R\right]. \tag{6}$$

The $8 \times 8$ matrix $\check{\kappa}$ describes the polarization of the magnetic interface separating the nanowire and the normal metal electrode and is aligned along the **z**-axis herein. The interface parameters

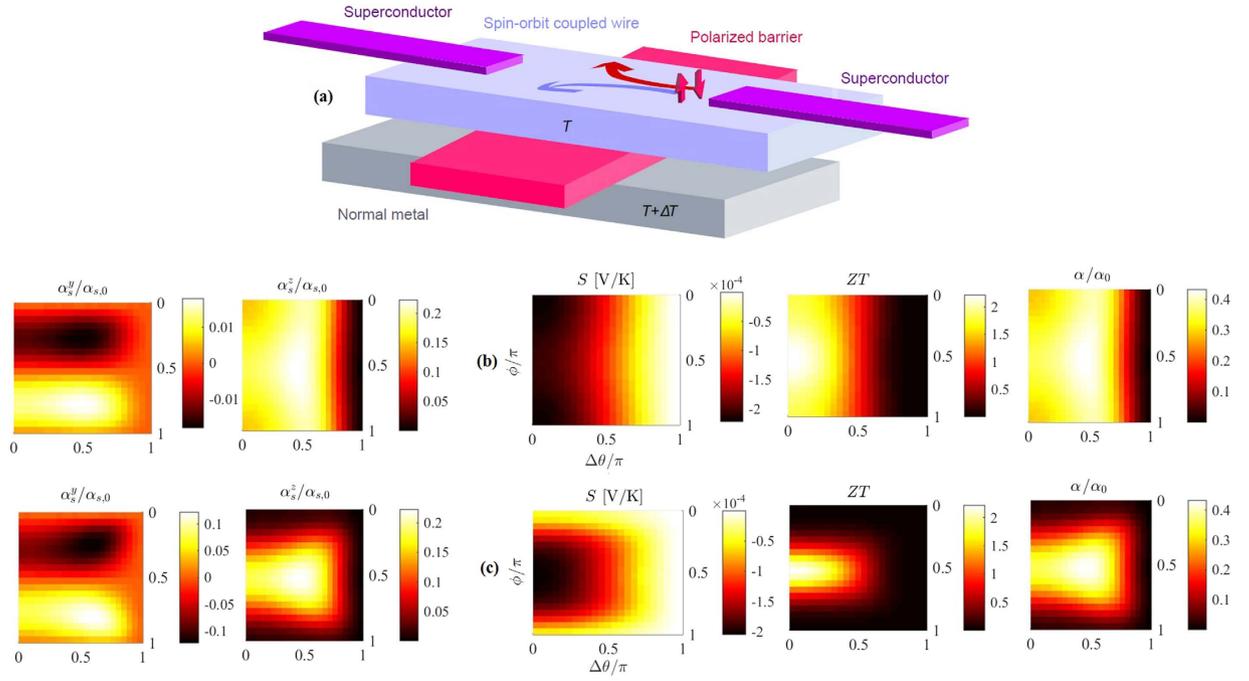

**Figure 3. Setup and thermoelectric effects for a superconducting structure incorporating a nanowire with Rashba spin-orbit coupling.** (**a**) Schematic of the proposed setup for observation of thermoelectric effects. Tunneling occurs from the center of a semiconductor nanowire into a normal metal electrode via a ferromagnetic insulator. The nanowire is connected to two superconducting reservoirs via tunneling barriers. The semiconductor nanowire is strongly spin-orbit coupled and exposed to a weak magnetic exchange field $h = 0.5\Delta_0$. *Left panels*: Spin thermal coefficients $\alpha_s^y/\alpha_{s,0}$ (left column) and $\alpha_s^z/\alpha_{s,0}$ (right column). The nanowire/electrode tunneling polarization is $P = 0$. The spin-orbit field strength is (**b**) $\beta L = 1$ and (**c**) $\beta L = 3$. *Right panels*: Seebeck coefficient $\mathcal{S}$ (left column), thermoelectric figure of merit $ZT$ (middle column), and thermoelectric coefficient $\alpha/\alpha_0$ (right column). The polarization for semiconductor/normal metal tunneling is $P = 97\%$.

$$G_0 = G_q \sum_{n=1}^{N} \tau_n (1 - \sqrt{1 - P_n^2}), \quad G_{MR} = G_q \sum_{n=1}^{N} \tau_n P_n,$$

$$G_1 = \sum_{n=1}^{N} \tau_n (1 + \sqrt{1 - P_n^2}), \quad G_\varphi = 2 G_q \sum_{n=1}^{N} \theta_n$$

(7)

describe interface resistance, barrier polarization and spin-dependent phase shifts occurring due to scattering at the interface. Barrier transparency is in the tunnelling limit, $G_q = e^2/h$ is the conductance quantum, $\tau_n$ the interface resistance and $P_n$ the polarization of transport channel $n$. We consider channel-independent scattering matrices where $\tau_n = \tau$ and $P_n = P$.

Thermoelectric effects arise in the geometries considered upon application of external voltage and temperature biases. Combining Eq. (6) with Eqns. (1) allows for calculation of the thermoelectric effects after performing a Taylor expansion in voltage and temperature to linear order for each current type. The Green function matrix to the left of the interface barrier describes the superconducting correlations induced in the nanowire while $\hat{g}_R^R = \hat{\rho}_3$ represents the normal metal electrode. The Green function matrices only depend on voltage and temperature via the distribution function matrices $\hat{h}_j$. The resulting thermoelectric coefficients are grouped together in a $2 \times 2$ Onsager matrix for linear response[55],

$$\begin{pmatrix} I_q \\ \dot{Q} \end{pmatrix} = \begin{pmatrix} L_{11} & L_{12} \\ L_{21} & L_{22} \end{pmatrix} \begin{pmatrix} \Delta V \\ \Delta T \end{pmatrix}.$$

(8)

Thermoelectric phenomena are commonly described using the Seebeck coefficient $\mathcal{S}$ and thermoelectric figure of merit $ZT$, defined by[56,57]

$$\mathcal{S} = -\frac{L_{12}}{L_{11}T}, \quad ZT = \left( \frac{L_{11} L_{22}}{L_{12}^2} - 1 \right)^{-1}.$$

(9)

The first step in computing the thermoelectric coefficients involves determining the unknown Green function matrices numerically. Herein, this is the Green function matrix within the nanowire, as we already know $\hat{g}_R^R = \hat{\rho}_3$.

The quasiclassical retarded Green function matrix on the left is determined in the middle of the central nanowire of the Josephson junction by solving the one-dimensional Usadel equation[58]

$$D\partial_x(\hat{g}^R \partial_x \hat{g}^R) = -i[E\hat{\rho}_3 + \hat{\Sigma}, \hat{g}^R],$$  (10)

where $\hat{\Sigma}$ is the self-energy term encompassing all material-specific properties such as magnetism and superconductivity. The subscript 'L' was omitted for brevity of notation. The Usadel equation (10) contains several $4 \times 4$ matrices and becomes cumbersome to solve for complex systems. Therefore, a Riccati parametrization[59] is performed to express $\hat{g}^R$ in terms of $2 \times 2$ $\underline{\gamma}$-matrices according to

$$\hat{g}^R = \begin{pmatrix} \underline{N}(1 + \underline{\gamma}\underline{\widetilde{\gamma}}) & 2\underline{N}\underline{\gamma} \\ -2\underline{\widetilde{N}}\underline{\widetilde{\gamma}} & -\underline{\widetilde{N}}(1 + \underline{\widetilde{\gamma}}\underline{\gamma}) \end{pmatrix},$$  (11)

where $\underline{N} = (1 - \underline{\gamma}\underline{\widetilde{\gamma}})^{-1}$ and $\underline{\widetilde{N}} = (1 - \underline{\widetilde{\gamma}}\underline{\gamma})^{-1}$. The Riccati-parametrized Usadel equation describing both normal metals, ferromagnetism and spin-orbit coupling is[36]

$$\begin{aligned}
D[\partial_x^2 \underline{\gamma} + 2\partial_x \underline{\gamma}\underline{\widetilde{N}}\underline{\widetilde{\gamma}}\partial_x \underline{\gamma}] &= -2iE\underline{\gamma} - ih(\sigma\underline{\gamma} - \underline{\gamma}\sigma^*) + D[AA\underline{\gamma} - \underline{\gamma}A^*A^*\\
&\quad + 2(A\underline{\gamma} + \underline{\gamma}A^*)\underline{\widetilde{N}}(A^* + \underline{\widetilde{\gamma}}A\underline{\gamma})]\\
&\quad + 2iD[\partial_x \underline{\gamma}\underline{\widetilde{N}}(A_x^* + \underline{\widetilde{\gamma}}A_x\underline{\gamma})\\
&\quad + (A_x + \underline{\gamma}A_x^*\underline{\widetilde{\gamma}})N\partial_x \underline{\gamma}],
\end{aligned}$$  (12)

where $h$ is the magnetic exchange field vector, $\sigma$ the Pauli vector and $A$ the spin-orbit field vector. The conversion reduces the amount of components the Usadel equation needs to be solved for, and can diminish the computational cost.

Boundary conditions describing the superconductor/nanowire interfaces well must be used in order to compute the $\underline{\gamma}$-matrices with satisfactory accuracy. The spin-active S/N interfaces of the S/N/S Josephson junction are described by[12,60]

$$\begin{aligned}
4L\zeta_L \partial_x \underline{\gamma}_L &= (1 - \underline{\gamma}_L\underline{\gamma}_R)\underline{N}_R(\underline{\gamma}_R - \underline{\gamma}_L) + 2iG_L^\varphi[\underline{\gamma}_L(\mathbf{m}\cdot\sigma^*) - (\mathbf{m}\cdot\sigma)\underline{\gamma}_L]\\
&\quad - 4\cosh(\Theta)G_{MR}[(\mathbf{m}\cdot\sigma)\underline{\gamma}_L + \underline{\gamma}_L(\mathbf{m}\cdot\sigma^*)]\\
4L\zeta_R \partial_x \underline{\gamma}_R &= (1 - \underline{\gamma}_R\underline{\gamma}_L)\underline{N}_L(\underline{\gamma}_R - \underline{\gamma}_L) - 2iG_R^\varphi[\underline{\gamma}_R(\mathbf{m}\cdot\sigma^*) - (\mathbf{m}\cdot\sigma)\underline{\gamma}_R]\\
&\quad + 4\cosh(\Theta)G_{MR}[(\mathbf{m}\cdot\sigma)\underline{\gamma}_R + \underline{\gamma}_R(\mathbf{m}\cdot\sigma^*)],
\end{aligned}$$  (13)

where $L$ is the length of the normal metal nanowire, $\zeta$ represents transparency at the superconductor/nanowire interfaces, $\Theta = \operatorname{atanh}\left(\frac{\Delta}{E+i\Gamma}\right)$, $\Gamma$ is the inelastic scattering energy scale, and $\Delta$ is the superconducting energy gap. The boundary conditions valid for the conical ferromagnet and the spin-orbit coupled semiconductor are the Kuprianov-Lukichev tunneling boundary conditions[61] modified for spin-orbit coupled materials[36]

$$\partial_x \underline{\gamma}_{L(R)} = \frac{1}{L\zeta_{L(R)}}(1 - \underline{\gamma}_{L(R)}\underline{\gamma}_{R(L)})\underline{N}_{R(L)}(\underline{\gamma}_R - \underline{\gamma}_L) + iA_x\underline{\gamma}_{L(R)} + i\underline{\gamma}_{L(R)}A_x^*.$$  (14)

## Results

**Quasiclassical thermoelectric coefficients.** The theoretical results were obtained based on a Josephson junction geometry, where a normal metal electrode is coupled to the central nanowire of the junction via an insulator polarized along the **z**-axis. The tunneling currents defined in Eqn. 1 were Taylor expanded to linear order w.r.t. voltage and temperature yielding the Onsager matrix

$$\begin{pmatrix} I_q \\ \dot{Q} \end{pmatrix} = \begin{pmatrix} G & \alpha \\ \alpha & G_Q \end{pmatrix}\begin{pmatrix} \Delta V \\ \Delta T/T \end{pmatrix}.$$  (15)

The thermoelectric coefficient, conductance coefficient and thermal conductance coefficient are given by

$$\alpha = \frac{G_T}{e}\int_{-\infty}^{\infty} \frac{EdE}{4k_BT \cosh^2\left(\frac{E}{2k_BT}\right)} \operatorname{Tr}\{\operatorname{Re}\{\hat{g}_L^R + P\hat{\sigma}_z\hat{g}_L^R\}\}$$  (16)

$$G = G_T \int_{-\infty}^{\infty} \frac{dE}{4k_BT \cosh^2\left(\frac{E}{2k_BT}\right)} \operatorname{Tr}\{\operatorname{Re}\{\hat{\rho}_3\hat{g}_L^R + P\hat{\rho}_3\hat{\sigma}_z\hat{g}_L^R\}\}$$  (17)

$$G_Q = \frac{G_\tau}{e^2} \int_{-\infty}^{\infty} \frac{E^2 dE}{4k_B T \cosh^2\left(\frac{E}{2k_B T}\right)} \operatorname{Tr}\{\operatorname{Re}\{\hat{\rho}_3 \hat{\mathbf{g}}_L^R + P\hat{\rho}_3 \hat{\sigma}_z \hat{\mathbf{g}}_L^R\}\}, \tag{18}$$

where $G_\tau = G_q N \tau$. The coefficients describe thermoelectric tunneling of charge and heat in superconducting hybrid systems with arbitrary spin-dependent magnetic textures and fields. In the limiting case of uniaxially aligned fields the expressions reduce to previous results in the literature[27]. The coefficients can be derived without assuming a normal metal electrode. The resulting expressions are more general, but also much more complex, and are valid whenever the previously mentioned constraints upon $\mu_L$, $\mu_R$, $\hat{h}_L$ and $\hat{h}_R$ are fulfilled. See Methods for further details.

The most notable analytical result of this work is obtained upon Taylor expanding the direction-dependent spin current with respect to the temperature, allowing the spin current to be expressed as $I_s^\nu = \alpha_s^\nu \frac{\Delta T}{T}$ when there is no applied voltage bias. Pure thermal spin currents[27,29] are predicted to arise in each spatial direction as a direct result of tunneling through the barrier between the normal metal electrode and the central nanowire of the Josephson junction. The expressions for the thermal spin coefficients $\alpha_s^\nu$ are

$$\alpha_s^x = \frac{G_\tau \hbar}{2e^2} \int_{-\infty}^{\infty} \frac{E dE}{4k_B T \cosh^2\left(\frac{E}{2k_B T}\right)} \sqrt{1-P^2} \operatorname{Tr}\{\operatorname{Re}\{\hat{\sigma}_x \hat{\mathbf{g}}_L^R\}\} \tag{19}$$

$$\alpha_s^y = \frac{G_\tau \hbar}{2e^2} \int_{-\infty}^{\infty} \frac{E dE}{4k_B T \cosh^2\left(\frac{E}{2k_B T}\right)} \sqrt{1-P^2} \operatorname{Tr}\{\operatorname{Re}\{\hat{\sigma}_y \hat{\mathbf{g}}_L^R\}\} \tag{20}$$

$$\alpha_s^z = \frac{G_\tau \hbar}{2e^2} \int_{-\infty}^{\infty} \frac{E dE}{4k_B T \cosh^2\left(\frac{E}{2k_B T}\right)} \operatorname{Tr}\{\operatorname{Re}\{\hat{\sigma}_z \hat{\mathbf{g}}_L^R + P \hat{\mathbf{g}}_L^R\}\}. \tag{21}$$

$\operatorname{Tr}\{\hat{\mathbf{g}}_L^R\}$ disappears in the quasiclassical approximation due to the restriction of charge neutrality. Accordingly, $\alpha_s^z$ is independent of barrier polarization. This is consistent with previous observations[27,29]. It is, however, important to note that the expressions are only valid when the quasiclassical approximation holds. The corresponding spin conductance coefficients are

$$G_s^x = \frac{G_\tau \hbar}{2e} \int_{-\infty}^{\infty} \frac{dE}{4k_B T \cosh^2\left(\frac{E}{2k_B T}\right)} \sqrt{1-P^2} \operatorname{Tr}\{\operatorname{Re}\{\hat{\rho}_3 \hat{\sigma}_x \hat{\mathbf{g}}_L^R\}\} \tag{22}$$

$$G_s^y = \frac{G_\tau \hbar}{2e} \int_{-\infty}^{\infty} \frac{dE}{4k_B T \cosh^2\left(\frac{E}{2k_B T}\right)} \sqrt{1-P^2} \operatorname{Tr}\{\operatorname{Re}\{\hat{\rho}_3 \hat{\sigma}_y \hat{\mathbf{g}}_L^R\}\} \tag{23}$$

$$G_s^z = \frac{G_\tau \hbar}{2e} \int_{-\infty}^{\infty} \frac{dE}{4k_B T \cosh^2\left(\frac{E}{2k_B T}\right)} \operatorname{Tr}\{\operatorname{Re}\{\hat{\rho}_3 \hat{\sigma}_z \hat{\mathbf{g}}_L^R + P \hat{\rho}_3 \hat{\mathbf{g}}_L^R\}\}, \tag{24}$$

which determine the voltage-driven spin current $I_s^\nu = G_s^\nu \Delta V$ in the absence of a temperature gradient. The expressions for the thermal spin and spin conductance coefficients presented above are a new result introduced herein, and together describe the system dynamics causing the spin Seebeck effect. The thermal spin coefficients demonstrate the possibility of generating thermal spin currents polarized along different spatial directions, depending on the spin-dependent fields within the materials being studied. The barrier for thermoelectric tunneling is defined to be polarized along the **z**-axis, explaining the prefactor $\sqrt{1-P^2}$ in front of the **x**- and **y**-directional coefficients $\alpha_s^x$ and $\alpha_s^y$. When the barrier is fully polarized along the **z**-axis, the spin Seebeck effect is suppressed in the other two directions.

In the next two sections, the new thermoelectric coefficients will be applied to different material systems in order to theoretically quantify the resulting thermoelectric effects. The Usadel equation must first be solved numerically in the middle of the nanowire followed by numerical integration to obtain the thermoelectric coefficients. Solving the Usadel equation only at one specific point in space limits the accuracy of the calculated thermoelectric effects, but using a narrow metal electrode should remedy the problem. For all the calculations presented herein we have used $L = 15$ nm as the nanowire length, $\zeta = 4$ to specify superconductor/nanowire interface transparency in the tunneling limit, $\Gamma = 0.005\Delta_0$ to represent inelastic scattering, $T = 0.2T_{c,0}$ for the temperature, $\xi = 30$ nm for the superconducting coherence length and $\Delta_0 = 1$ meV for the superconducting energy gap. The superconducting coherence length is chosen to represent Nb with $\xi_0 = 38$ nm, $\Delta_0 = 1.5$ meV and a superconducting critical temperature of $T_c = 9.5$ K, where the last is highest of all the elemental superconductors[62].

**Thermoelectric figure of merit and Seebeck coefficient.** The thermoelectric figure of merit $ZT$ and Seebeck coefficient $\mathcal{S}$ are studied for three different device scenarios. $ZT$ and $\mathcal{S}$ are defined in Eqn. 9 and we use the coefficients derived in Eqns. 15–18. The Seebeck coefficient is maximized when the nanowire/electrode interface polarization is as large as possible. We defined the interface polarization to be $P = 97\%$, consistent with the polarization of the ferromagnetic insulator GdN at 3 K[63]. All three geometries studied are derived from the S/X/S Josephson junction where thermoelectric phenomena arise from tunneling between the nanowire X and a normal metal electrode. As previously mentioned, the nanowire is either a normal metal with magnetic interfaces to the superconductors, a conical ferromagnet or a semiconductor containing spin-orbit coupling.

Figure 1(a) shows a graphical representation of the proposed material setup containing a conical ferromagnet as the central nanowire of the Josephson junction. The magnetic texture is spatially varying, and no external magnetic fields need be applied for thermoelectric phenomena to arise. One of the best known conical ferromagnets is the material holmium (Ho)[64,65] below 19 K[66]. The conical ferromagnet is described by the complete Usadel equation in Eq. 12 with a spin-orbit field $A = (0, 0, 0)$ and magnetic field vector $h = (h_x, h_y, h_z)$ defined by

$$h_x = h\cos(\varphi), \quad h_y = h\sin(\varphi)\sin\left(\frac{\theta x}{a}\right), \quad h_z = h\sin(\varphi)\cos\left(\frac{\theta x}{a}\right). \quad (25)$$

The material-specific constants $\varphi$, $a$ and $\theta$ are chosen as $\varphi = \frac{4\pi}{9}$, $\theta = \pi/6$ and $a = 0.526$ nm to represent Ho[67,68]. As the magnetic exchange field in Ho has been reported to have different sizes in various experiments[17,69–71], we here consider thermoelectric response over a range of field strengths.

The Seebeck coefficient and the thermoelectric figure of merit arising due to tunneling from the conical ferromagnet are quite small, only reaching $S = -0.06$ mV/K and $ZT = 0.025$, and are therefore not shown. This is substantially smaller than what is obtainable in Zeeman-split superconducting hybrids. The thermoelectric coefficient, on the other hand, approaches $\alpha/\alpha_0 = 0.1$ in the best case scenario where the exchange field in the conical ferromagnet is $h \sim 3\Delta_0$. This is of the same order of magnitude as the thermoelectric coefficient governing thermal charge and spin transport in Zeeman-split superconducting hybrids. The qualitative behavior of $\alpha/\alpha_0$ is equal to that of the thermal spin coefficient along the **z**-axis, $\alpha_s^z/\alpha_{s,0}$ (Fig. 1(d)), and is not included here. The thermoelectric phenomena induced by the conical ferromagnet vary with the ferromagnetic exchange field $h$ and the superconducting phase difference $\Delta\theta$. We have defined $\alpha_0 = G_\tau \Delta_0/e$ and $\alpha_{s,0} = G_\tau \hbar \Delta_0/e^2$.

Figure 2(a) shows the proposed device setup for the superconducting hybrid incorporating spin-active superconductor/nanowire interfaces. The nanowire is a non-magnetic normal metal but the S/N interfaces are occupied by thin, weakly polarized ferromagnetic insulators. These spin-active interfaces are described by Cottet's boundary conditions (Eq. 13). The right S/N interface is aligned along the **z**-axis, defined by $m_R = \sigma_z$, while the alignment of the left interface can be varied in the **yz**-plane according to $m_L = \cos(\phi)\sigma_z + \sin(\phi)\sigma_y$. The thermoelectric effects arising through tunneling from the nanowire to the electrode are presented as functions of the superconducting phase difference $\Delta\theta$ and the alignment angle $\phi$ of the left S/N interface in the **yz**-plane. The magnetic field at this interface is aligned along the **z**-axis when $\phi = 0$ and along the **y**-axis when $\phi = \pi/2$. The remaining interface parameters for S/N tunneling are $G_{MR} = 0.1$ indicating weak interface polarization and $G_\varphi = 0.5$ or 1.05 representing spin-dependent phase shifts resulting from scattering at the interfaces.

The Seebeck coefficient, the thermoelectric figure of merit and the thermoelectric coefficient in the case of the spin-active Josephson junction are shown in the bottom right panel of Fig. 2. The top row (b) of the panel shows $G_\varphi = 0.5$, and the bottom row (c) of the panel shows $G_\varphi = 1.05$. The resulting thermoelectric effects rival the magnitudes obtained in the Zeeman-split superconducting bilayer from ref. 27. By selectively tuning $\Delta\theta$ and $\phi$ we can maximally obtain $S = 0.2$ mV/K, $ZT = 2$ and $\alpha/\alpha_0 = -0.4$. The magnetic fields necessary to reorient the magnetic S/N interfaces are much weaker than those needed to induce a strong Zeeman exchange field comparable in magnitude to the superconducting gap $\Delta_0$. When $G_\varphi = 0.5$, the maximum thermoelectric effects are obtained when both S/N interfaces are aligned in parallel along the **z**-axis, as seen in Fig. 2(b). Upon increasing $G_\varphi$ to 1.05 in Fig. 2(c), interface scattering causes a larger degree of spin-dependent phase shifts, moving the alignment angles maximizing the thermoelectric phenomena described by $S$, $ZT$ and $\alpha$ closer to the **y**-axis and $\phi = \pi/2$.

Figure 3(a) shows the third and last scenario where a doped spin-orbit coupled semiconductor constitutes the central nanowire of the Josephson junction. The primary reason for employing a semiconductor for this purpose is the possibility of a large Landé g-factor, allowing for enhanced spin response upon the application of a magnetic field[72]. Reportedly, the Landé g-factor takes the value $g \approx 2$ in superconducting Al[73], but can reach $g \approx 10$–20 in spin-orbit coupled InAs nanowires[74,75]. The external fields needed to induce a significant particle-hole asymmetry and generate thermoelectric phenomena in spin-orbit coupled superconducting hybrids are therefore much smaller than the aforementioned Zeeman-field of $\sim 1$ T. Within this work we only study Rashba spin-orbit coupling, and the spin-orbit field is defined as $A = (A_x, 0, 0)$ where

$$A_x = \beta\sin(\phi)\sigma_z - \beta\cos(\phi)\sigma_y. \quad (26)$$

$\beta$ determines the spin-orbit field strength and $\phi$ is the field alignment angle in the **yz**-plane. The magnetic field is applied along the **z**-axis, so $h = (0, 0, h)$. Within this framework $\phi = 0$ indicates field alignment along the $-$**y**-axis, $\phi = \pi/2$ a spin-orbit field along the $+$**z**-axis and $\phi = \pi$ field alignment along the $+$**y**-axis. In practice, the variation of $\phi$ can be achieved by either rotating the sample itself or rotating the external field as both of these procedures are fully equivalent[76].

The bottom right panel of Fig. 3 shows the thermoelectric effects arising in the spin-orbit coupled Josephson junction geometry. They are comparable in size to the spin-active case depicted in Fig. 2, and therefore also to the high field Zeeman-split bilayers. Seebeck coefficients approaching $S = 0.2$ mV/K, thermoelectric figures of merit $ZT = 2$ and thermoelectric coefficients $\alpha/\alpha_0 = 0.4$ seem to be obtainable in such a configuration. The material parameters studied include an externally applied magnetic exchange field $h = 0.5\Delta_0$ and spin-orbit coupling strengths (b) $\beta L = 1$ and (c) $\beta L = 3$. Changing the spin-orbit field strength is seen to affect how the thermoelectric coefficients vary with the field alignment angle $\phi$ in the **yz**-plane. When $\beta L = 1$ a large change in the field alignment has very little effect on the size of $\alpha$, $\mathcal{S}$ and $ZT$, while the superconducting phase difference determines whether thermoelectric phenomena exist or not. This is even more pronounced when the spin-orbit field is weaker and $\beta L = 0.1$, but this is not shown herein. $\Delta\theta = \pi$ suppresses superconducting order within the nanowire, effectively preventing thermoelectric effects and causing $\mathcal{S} = ZT = \alpha = 0$. The maximum values of the different thermoelectric coefficients are not altered significantly as the spin-orbit field strength is increased. However, increasing the spin-orbit field to $\beta L = 3$ (Fig. 3(c)) makes tuning the field alignment angle correctly crucial. The underlying physical reason for this is the large anisotropy in the depairing energy penalty of the spin-triplet Cooper pairs induced in the nanowire region[36], which is controlled via the field orientation. The maximum values for the thermoelectric coefficient $\alpha$, the Seebeck coefficient $\mathcal{S}$ and the thermoelectric figure of merit $ZT$ are found when the spin-orbit field alignment angle equals $\phi = \pi/2$, as seen in the right panel of Fig. 3(c). At this angle, the field is aligned along the **z**-axis, in the same direction as the magnetic exchange field $\mathbf{h} = (0, 0, h)$.

**Spin Seebeck effect.** The spin Seebeck effect will here be studied in further detail for superconducting hybrids with magnetic texturing. Depending on the spin fields within the material systems chosen, observation of pure thermal spin currents which are independent of the interface polarization is theoretically possible. When studying the spin Seebeck effect, we consider the case of $P = 0$ for the tunneling barrier between the Josephson junction and the normal metal electrode. This is done in order to maximize the spin Seebeck effect along the **x**- and **y**-axes as the corresponding thermal spin coefficients $\alpha_s^\nu$ are proportional to $\sqrt{1 - P^2}$. The tunneling barrier is defined to be polarized along the **z**-axis, causing $\alpha_s^x$ and $\alpha_s^y$ to diminish with increasing polarization and disappear entirely when $P = 100\%$.

The conical S/F/S Josephson junction is the only configuration for which the thermal spin current along the **x**-axis is dominant. The thermal spin coefficients arising in this scenario are depicted in Fig. 1(b–d). The usual pair-breaking effect of the ferromagnetic exchange field is less pronounced when conical magnetic texturing is present, even though the conventional thermoelectric effects quantified by $\mathcal{S}$ and $ZT$ are rather small. The quantitative behavior of the Seebeck coefficient and thermoelectric figure of merit is directly related to the evolution and size of $\alpha_s^z$ due to its proportionality to the thermoelectric coefficient $\alpha$. The lack of significant thermally driven electric currents does, however, not prevent prominent thermal spin currents from traversing the system.

The thermal spin coefficient $\alpha_s^x$ is vanishingly small in the last two material systems, an effect which is directly related to spin-dependent field alignment within the **yz**-plane. A graphical representation of $\alpha_s^x$ is therefore not included in this work for these systems. The thermal spin currents in the other two directions are much larger in both cases, and are shown in the bottom left panels of Figs 2 and 3.

A notable feature when considering the spin-active Josephson junction, and comparing Fig. 2(b,c), is the increase in $\alpha_s^y$ when increasing $G_\varphi$ from 0.55 to 1.05. Increasing spin-dependent phase shifts at the interface seems to force quasiparticle spins to align along the **y**-axis as opposed to the **z**-axis. The field alignment angle causing maximal thermal spin currents is also affected by changing $G_\varphi$, an effect which is more noticeable for even larger values of $G_\varphi$ than depicted herein. The thermal spin currents, along with $\mathcal{S}$ and $ZT$, seem to disappear as the interface field alignment angle reaches $\phi = \pi$. At this angle the magnetic fields at the S/N interfaces are aligned in exactly opposite directions. All thermoelectric phenomena become vanishingly small in this limit. This may be understood physically from the suppression of the spin-triplet Cooper pairs in this configuration as the net exchange field is averaged out in the center of the nanowire. When the triplet proximity effect vanishes, so does the spin-dependent particle-hole asymmetry of the system.

The thermal spin coefficient along the **z**-axis behaves in the same manner as the thermoelectric coefficient $\alpha$ when the spin-orbit coupled Josephson junction is considered. The maximum value of $\alpha_s^z$ is largely unaffected upon increasing the spin-orbit field strength, as can be seen when comparing Fig. 3(b,c). Increasing the spin-orbit field does, however, affect how rapidly the thermal spin coefficients change when the field alignment angle is varied. The behavior of $\alpha_s^y$ is fundamentally different, as this coefficient is sinusoidal in the field alignment angle. This sinusoidal shape is consistent as the field strength is increased, while the the quantitative change is more pronounced. In contrast to the thermal spin current generated along the axis of the magnetic exchange field, thermal $I_s^y$ requires a larger spin-orbit field to reach a substantial size, in this case at least $\beta L = 3$. The Rashba coefficient $\beta$ was normalized with respect to $\hbar^2/L$. Depending on the electron effective mass, the normalized Rashba coupling strength $\beta L = 3$ corresponds to a Rashba coefficient $\beta/m^* = 1.52 \times 10^{-11}$ eVm when $m^*$ equals the free electron mass, $m_0 = 9.11 \times 10^{-31}$ kg. This fits quite well with for instance the experimentally determined Rashba coefficient in InAlAs/InGaAs ($\sim 0.67 \times 10^{-11}$ eVm)[77]. The sinusoidal behavior of $\alpha_s^y$ seems to depend only upon the field alignment angle, with the thermal spin coefficient being positive when the spin-orbit field is aligned in the $+$**yz**-plane ($\phi \in [0.5\pi, \pi]$) and negative for angles within the plane between the $-$**y**- and $+$**z**-axes ($\phi \in [0, 0.5\pi]$). The direction of the thermal spin current along the **y**-axis is thus controllable simply by altering the orientation of the weak external magnetic field.

A prominent feature occurring for all the thermoelectric coefficients studied herein is the disappearance of the thermoelectric effects as the superconducting phase difference reaches $\Delta\theta = \pi$. Thermoelectric effects at this phase difference would indicate the existence of asymmetries in the density of states in the middle of the Josephson junction central nanowire. Superconducting order is known to be suppressed in most Josephson junctions at this phase difference. However, recent studies have emphasized the presence of such superconducting order when $\Delta\theta = \pi$ in Josephson junctions containing strong spin-orbit coupling[37]. Thermoelectric phenomena arising when $\Delta\theta = \pi$ were therefore expected to some degree, particularly in the case of the spin-orbit coupled Josephson junction. The presence of such asymmetries for the specified phase difference were discovered in the density of states in several of the structures considered, but at magnitudes much too low to result in detectable thermoelectric or thermal spin currents. The presence of these asymmetries in all three spatial directions indicates the possibility of large thermoelectric effects arising even when the superconducting phase difference is $\Delta\theta = \pi$, but for different choices of material properties and specific parameters.

The spin-Seebeck coefficient was calculated in the same manner as the Seebeck coefficient. The spin-Seebeck coefficient depends on spatial alignment in the same manner as the spin current, and is defined as

$$\mathcal{S}_s^\nu = -\frac{\alpha_s^\nu}{G_s^\nu T}. \tag{27}$$

The unit of the spin-Seebeck coefficient is once again V/K, and the coefficients $\mathcal{S}_s^\nu$ are therefore directly comparable to $\mathcal{S}$. The maximum values of the spin-Seebeck coefficients are almost equally large as the regular Seebeck coefficient in some directions. Notable spin Seebeck coefficients are $\mathcal{S}_{s,\max}^x = -1.25 \times 10^{-5}$ V/K in the case of the spin-active Josephson junction with $G_\varphi = 1.05$ when $P = 0$ and $P = 97\%$, and $\mathcal{S}_{s,\max}^z = 1 \times 10^{-4}$ V/K for the same material system with tunneling polarization $P = 97\%$. The spin-orbit coupled Josephson junction with tunneling polarization $P = 97\%$ is capable of producing spin-Seebeck coefficients $\mathcal{S}_{s,\max}^z \approx -2.2 \times 10^{-4}$ V/K when $\beta L = 1$ and $\beta L = 3$. This is practically identical to $\mathcal{S}$ for the same theoretical scenarios.

**Concluding remarks.** A framework for calculating thermoelectric coefficients in systems with arbitrary spin-dependent field alignment was derived and applied to theoretical device geometries. The results presented herein demonstrate the effect of spin-active interfaces, textured ferromagnetism and Rashba spin-orbit interactions on thermoelectric phenomena in superconducting hybrids. The spin-dependent fields present in such materials are capable of generating large thermoelectric and spin Seebeck effects even in the absence of strong external magnetic fields. Small external fields do need to be applied to generate thermal electric and spin currents exiting the spin-orbit coupled Josephson nanowires, but they should be much smaller than the ~1 T fields necessary for Zeeman splitting of the superconducting density of states considered in previous works. Thermoelectric phenomena comparable to those arising in Zeeman-split geometries were predicted, both of the conventional kind and also including pure thermal spin currents polarized in different directions.

## Methods

The Onsager response matrix and quasiclassical thermoelectric coefficients presented in the main text are only valid when the electrode coupled to the nanowire is a normal metal. More general expressions were initially derived but subsequently simplified to the ones presented above. The complete thermoelectric coefficients for a random choice of materials for both the nanowire and the electrode are

$$\begin{aligned} G &= \frac{G_\tau}{4} \int_{-\infty}^{\infty} \frac{dE}{4k_B T \cosh^2\left(\frac{E}{2k_B T}\right)} \operatorname{Tr}\{(1 + \sqrt{1-P^2})\operatorname{Re}\{\hat{g}_L \hat{g}_R + \hat{g}_L \hat{g}_R^\dagger\} \\ &\quad + P \operatorname{Re}\{\hat{\sigma}_z \hat{g}_L \hat{g}_R + \hat{\sigma}_z \hat{g}_R \hat{g}_L + \hat{\sigma}_z \hat{g}_L \hat{g}_R^\dagger + \hat{\sigma}_z \hat{g}_R^\dagger \hat{g}_L\} \\ &\quad + (1 - \sqrt{1-P^2})\operatorname{Re}\{\hat{\sigma}_z \hat{g}_L \hat{\sigma}_z \hat{g}_R + \hat{\sigma}_z \hat{g}_L \hat{\sigma}_z \hat{g}_R^\dagger\}\}, \end{aligned} \tag{28}$$

$$\begin{aligned} \alpha_1 &= T\frac{dI_q}{dT_L} = \frac{G_\tau}{4e} \int_{-\infty}^{\infty} \frac{EdE}{4k_B T \cosh^2\left(\frac{E}{2k_B T}\right)} \operatorname{Tr}\{(1 + \sqrt{1-P^2})\operatorname{Re}\{\hat{\rho}_3 \hat{g}_R \hat{g}_L + \hat{\rho}_3 \hat{g}_R^\dagger \hat{g}_L\} \\ &\quad + P \operatorname{Re}\{\hat{\rho}_3 \hat{\sigma}_z \hat{g}_R \hat{g}_L + \hat{\rho}_3 \hat{\sigma}_z \hat{g}_R^\dagger \hat{g}_L + \hat{\rho}_3 \hat{g}_R \hat{\sigma}_z \hat{g}_L + \hat{\rho}_3 \hat{g}_R^\dagger \hat{\sigma}_z \hat{g}_L\} \\ &\quad + (1 - \sqrt{1-P^2})\operatorname{Re}\{\hat{\rho}_3 \hat{\sigma}_z \hat{g}_R \hat{\sigma}_z \hat{g}_L + \hat{\rho}_3 \hat{\sigma}_z \hat{g}_R^\dagger \hat{\sigma}_z \hat{g}_L\}\}, \end{aligned} \tag{29}$$

$$\begin{aligned} \alpha_2 &= T\frac{dI_q}{dT_R} = -\frac{G_\tau}{4e} \int_{-\infty}^{\infty} \frac{EdE}{4k_B T \cosh^2\left(\frac{E}{2k_B T}\right)} \operatorname{Tr}\{(1 + \sqrt{1-P^2})\operatorname{Re}\{\hat{\rho}_3 \hat{g}_L \hat{g}_R + \hat{\rho}_3 \hat{g}_R^\dagger \hat{g}_L\} \\ &\quad + P\operatorname{Re}\{\hat{\rho}_3 \hat{\sigma}_z \hat{g}_L \hat{g}_R + \hat{\rho}_3 \hat{\sigma}_z \hat{g}_R^\dagger \hat{g}_L + \hat{\rho}_3 \hat{g}_L \hat{\sigma}_z \hat{g}_R + \hat{\rho}_3 \hat{g}_R^\dagger \hat{\sigma}_z \hat{g}_L\} \\ &\quad + (1 - \sqrt{1-P^2})\operatorname{Re}\{\hat{\rho}_3 \hat{\sigma}_z \hat{g}_L \hat{\sigma}_z \hat{g}_R + \hat{\rho}_3 \hat{\sigma}_z \hat{g}_R^\dagger \hat{\sigma}_z \hat{g}_L\}\}, \end{aligned} \tag{30}$$

and

$$\begin{aligned} G_Q &= \frac{G_\tau}{4e^2} \int_{-\infty}^{\infty} \frac{E^2 dE}{4k_B T \cosh^2\left(\frac{E}{2k_B T}\right)} \operatorname{Tr}\{(1 + \sqrt{1-P^2})\operatorname{Re}\{\hat{g}_R \hat{g}_L + \hat{g}_L \hat{\rho}_3 \hat{g}_R^\dagger \hat{\rho}_3\} \\ &\quad + P \operatorname{Re}\{\hat{\sigma}_z \hat{g}_R \hat{g}_L + \hat{g}_R \hat{\sigma}_z \hat{g}_L + \hat{\rho}_3 \hat{g}_L \hat{\rho}_3 \hat{g}_R^\dagger \hat{\sigma}_z + \hat{g}_L \hat{\sigma}_z \hat{\rho}_3 \hat{g}_R^\dagger \hat{\rho}_3\} \\ &\quad + (1 - \sqrt{1-P^2})\operatorname{Re}\{\hat{\sigma}_z \hat{g}_R \hat{\sigma}_z \hat{g}_L + \hat{g}_L \hat{\sigma}_z \hat{\rho}_3 \hat{g}_R^\dagger \hat{\sigma}_z \hat{\rho}_3\}\}. \end{aligned} \tag{31}$$

Once again, $G_\tau = G_q N \tau$, $G_q = e^2/h$ is the conductance quantum, $N$ is the number of tunneling channels and $\tau$ is the nanowire/electrode interface transparency. The thermal spin coefficient can be written as

$$\alpha^{\nu}_{s,1} = T\frac{dI^{\nu}_s}{dT_L} = \frac{G_{\tau}}{16e^2}\int_{-\infty}^{\infty}\frac{EdE}{4k_BT\cosh^2\left(\frac{E}{2k_BT}\right)}\text{Tr}\{\hat{\rho}_3\hat{\tau}_{\nu}[(1+\sqrt{1-P^2})(\hat{g}_L\hat{\rho}_3\hat{g}_R^{\dagger}\hat{\rho}_3$$
$$+\hat{\rho}_3\hat{g}_L^{\dagger}\hat{g}_R^{\dagger}\hat{\rho}_3 + \hat{g}_R\hat{g}_L + \hat{g}_R\hat{\rho}_3\hat{g}_L^{\dagger}\hat{\rho}_3) + P(\hat{\sigma}_z\hat{g}_L\hat{\rho}_3\hat{g}_R^{\dagger}\hat{\rho}_3 + \hat{\sigma}_z\hat{\rho}_3\hat{g}_L^{\dagger}\hat{g}_R^{\dagger}\hat{\rho}_3$$
$$+\hat{g}_L\hat{\sigma}_z\hat{\rho}_3\hat{g}_R^{\dagger}\hat{\rho}_3 + \hat{\rho}_3\hat{g}_L^{\dagger}\hat{\sigma}_z\hat{g}_R^{\dagger}\hat{\rho}_3 + \hat{g}_R\hat{\sigma}_z\hat{g}_L + \hat{g}_R\hat{\sigma}_z\hat{\rho}_3\hat{g}_L^{\dagger}\hat{\rho}_3 + \hat{g}_R\hat{g}_L\hat{\sigma}_z$$
$$+\hat{g}_R\hat{\rho}_3\hat{g}_L^{\dagger}\hat{\rho}_3\hat{\sigma}_z) + (1-\sqrt{1-P^2})(\hat{\sigma}_z\hat{g}_L\hat{\sigma}_z\hat{\rho}_3\hat{g}_R^{\dagger}\hat{\rho}_3$$
$$+\hat{\sigma}_z\hat{\rho}_3\hat{g}_L^{\dagger}\hat{\sigma}_z\hat{g}_R^{\dagger}\hat{\rho}_3 + \hat{g}_R\hat{\sigma}_z\hat{g}_L\hat{\sigma}_z + \hat{g}_R\hat{\rho}_3\hat{\sigma}_z\hat{g}_L^{\dagger}\hat{\rho}_3\hat{\sigma}_z)]\}$$
(32)

and

$$\alpha^{\nu}_{s,2} = T\frac{dI^{\nu}_s}{dT_R} = -\frac{G_{\tau}}{16e^2}\int_{-\infty}^{\infty}\frac{EdE}{4k_BT\cosh^2\left(\frac{E}{2k_BT}\right)}\text{Tr}\{\hat{\rho}_3\hat{\tau}_{\nu}[(1+\sqrt{1-P^2})$$
$$\times(\hat{g}_L\hat{g}_R + \hat{g}_L\hat{\rho}_3\hat{g}_R^{\dagger}\hat{\rho}_3 + \hat{g}_R\hat{\rho}_3\hat{g}_L^{\dagger}\hat{\rho}_3 + \hat{\rho}_3\hat{g}_R^{\dagger}\hat{g}_L^{\dagger}\hat{\rho}_3) + P(\hat{\sigma}_z\hat{g}_L\hat{g}_R + \hat{\sigma}_z\hat{g}_L\hat{\rho}_3\hat{g}_R^{\dagger}\hat{\rho}_3$$
$$+\hat{g}_L\hat{\sigma}_z\hat{g}_R + \hat{g}_L\hat{\sigma}_z\hat{\rho}_3\hat{g}_R^{\dagger}\hat{\rho}_3 + \hat{g}_R\hat{\sigma}_z\hat{\rho}_3\hat{g}_L^{\dagger}\hat{\rho}_3 + \hat{\rho}_3\hat{g}_R^{\dagger}\hat{\sigma}_z\hat{g}_L^{\dagger}\hat{\rho}_3 + \hat{g}_R\hat{\rho}_3\hat{g}_L^{\dagger}\hat{\rho}_3\hat{\sigma}_z$$
$$+\hat{\rho}_3\hat{g}_R^{\dagger}\hat{g}_L^{\dagger}\hat{\rho}_3\hat{\sigma}_z) + (1-\sqrt{1-P^2})(\hat{\sigma}_z\hat{g}_L\hat{\sigma}_z\hat{g}_R + \hat{\sigma}_z\hat{g}_L\hat{\sigma}_z\hat{\rho}_3\hat{g}_R^{\dagger}\hat{\rho}_3$$
$$+\hat{g}_R\hat{\sigma}_z\hat{\rho}_3\hat{g}_L^{\dagger}\hat{\rho}_3\hat{\sigma}_z + \hat{\rho}_3\hat{g}_R^{\dagger}\hat{\sigma}_z\hat{g}_L^{\dagger}\hat{\rho}_3\hat{\sigma}_z) - iG_{\varphi}(\hat{\sigma}_z\hat{g}_R + \hat{\sigma}_z\hat{\rho}_3\hat{g}_R^{\dagger}\hat{\rho}_3$$
$$-\hat{g}_R\hat{\sigma}_z - \hat{\rho}_3\hat{g}_R^{\dagger}\hat{\rho}_3\hat{\sigma}_z)]\}.$$
(33)

The general thermoelectric coefficients rely on no assumptions regarding the nature of the materials as long as they comply with the restrictions $\mu_L = 0$, $\mu_R = eV_R$, $\hat{h}_L = \tanh\left(\frac{\beta_L E}{2}\right)\hat{\mathbf{1}}$ and $\hat{h}_R$ as defined in Eq. 4.

### Acknowledgements

J.L. acknowledges support from the Outstanding Academic Fellows programme at NTNU and the Norwegian Research Council Grant No. 216700 and No. 240806.

### Author Contributions

M.E.B. performed the analytical and numerical calculations with minor support from J.L. Both authors contributed to the writing and discussion of the manuscript.

### Additional Information

**Competing financial interests:** The authors declare no competing financial interests.

**How to cite this article**: Bathen, M. E. and Linder, J. Spin Seebeck effect and thermoelectric phenomena in superconducting hybrids with magnetic textures or spin-orbit coupling. *Sci. Rep.* **7**, 41409; doi: 10.1038/srep41409 (2017).

**Publisher's note:** Springer Nature remains neutral with regard to jurisdictional claims in published maps and institutional affiliations.